\documentclass{siamart220329}
\usepackage{amsmath,graphicx,amsfonts,bm,color}
\renewcommand{\textwidth}{6.5in}
\renewcommand{\textheight}{9in}

\newcommand{\reals} {\mathbb{R}}
\newcommand{\boldk} {\bm{k}}
\newcommand{\boldx} {\bm{x}}
\newcommand{\boldy} {\bm{y}}
\newcommand{\boldu} {\bm{u}}
\newcommand{\boldZ} {\bm{Z}}
\newcommand{\boldX} {\bm{X}}
\newcommand{\boldY} {\bm{Y}}

\newcommand{\Cap}{\mathrm{C} }
\newcommand{\Capalpha}{\Cap_\alpha}
\newcommand{\Rlaunch}{R_\text{launch}}
\newcommand{\Rescape}{R_\text{escape}}

\newcommand{\ball}{\mathbb{B}}
\newcommand{\sphere}{\mathbb{S}}

\ifpdf
  \DeclareGraphicsExtensions{.eps,.pdf,.png,.jpg}
\else
  \DeclareGraphicsExtensions{.eps}
\fi

\newsiamremark{remark}{Remark}
\newsiamremark{hypothesis}{Hypothesis}
\crefname{hypothesis}{Hypothesis}{Hypotheses}
\newsiamthm{claim}{Claim}

\headers{$\alpha$-capacity of general sets}{J. P. Nolan, D. J. Audus, and J. F. Douglas}

\title{ Computation of Riesz $\alpha$-capacity $\Capalpha$ of general sets in  $\reals^d$ using stable random
walks\thanks{Submitted to the editors \today.}}
\author{John P. Nolan\thanks{American University, Department of Mathematics and Statistics, Washington DC 20016 and
   Applied and Computational Mathematics Division, National Institute of Standards and Technology, Gaithersburg, Maryland 20899
  (\email{jpnolan@american.edu}).}
\and Debra J. Audus\thanks{Materials Science and Engineering Division,
   National Institute of Standards and Technology, Gaithersburg, Maryland 20899
   (\email{debra.audus@nist.gov}, \email{jack.douglas@nist.gov}).}
\and Jack F. Douglas\footnotemark[3]}

\usepackage{amsopn}

\begin{document}
% -------------------------------------------------------------------------

\maketitle

\begin{abstract}
A method for computing the Riesz $\alpha$-capacity, $0 < \alpha \le 2$, of a general set $K \subset \reals^d$ is given.
The method is based on simulations of isotropic $\alpha$-stable motion paths in $d$-dimensions.  The familiar
Walk-On-Spheres method, often utilized for simulating Brownian motion, is
modified to a novel Walk-In-Out-Balls method adapted for  modeling the stable path process
on the exterior of regions ``probed'' by this type of generalized random walk.  It accounts for the propensity of this
class of random walk to jump through boundaries because of the path discontinuity.
This method allows for the computationally efficient simulation of
hitting locations of stable paths launched from the exterior of probed sets.  Reliable methods of computing capacity from these locations are given,
 along with non-standard confidence intervals.
 Illustrative calculations are performed for representative types of sets K, where both $\alpha$ and $d$ are varied.
\end{abstract}


% -------------------------------------------------------------------------
\section{Introduction}

The capacity of a set $K \subset \reals^d$ is a measure of the ``size'' of the set,
and when combined with the volume of the set, this quantity gives information about the
shape of the set.  Capacity is related to many physical properties of an object, e.g.
the electrostatic capacity of charged objects, the influence of the shape of
reactive particles on the rate of diffusion-limited chemical reactions,
how objects diffuse in liquids,  how heat flows to and from an object, etc.
These and other applications are
discussed in Douglas et al. \cite{douglas:zhou:hubbard:1994}. Recent work has shown that the
shape of a biological cell is relevant to its functions, and Betancourt et al.
\cite{betancourt:etal:2018} use capacity as a metric for quantifying cell shapes.

Recently, there has been great interest in “generalized boundary value problems” involving
“fractional diffusion processes” related to Lévy stable processes. The connection between these
fractional differential equations and Lévy stable path processes arises from the fact that the
fractional Laplacian is the generating operator of Lévy stable processes. There has also been
interest and scientific activity in numerical computations aimed at obtaining
solutions of this type of generalized boundary value problem based on formal extensions of
classical probabilistic potential theory ideas developed for Brownian motion, whose
generator is the ordinary Laplacian operator, to solving fractional differential boundary value
problems. There are many unsolved problems in this active field of study, and we mention the
related  numerical study of Kyprianou et al. \cite{kyprianou:osojnik:shardlow:2018}, which addresses the solution of extensions
of classical interior boundary value problems, such as Poisson equation, where the sampling of
the well-known walk-on-sphere (WOS) methodology was utilized as in previous numerical
simulations based on ordinary random walks. Encouragingly, they found that it was possible to
obtain numerical results in good agreement with the limited known analytic results for stable
processes. In contrast, our paper is devoted to an extension of the solution of Laplace’s
equation on the exterior of a generally shaped region $K$ in $d$ dimensions, the exterior
Dirichlet problem. We are in particular interested in precisely estimating the Riesz alpha
capacity $\Capalpha$  in $d$ spatial dimensions, which describes how the harmonic solution to this
classical exterior boundary value problem decays at large distances from $K$. We find that the
treatment of this type of exterior problem unfortunately does not admit to the simple WOS
methodology because of the non-continuity of the Lévy path processes for $\alpha < 2$ so that we
must introduce new methods of accelerating the path sampling required for our exterior
boundary value problem simulations of $\Capalpha$.

The capacity is a shape functional that has numerous applications.  Methodologies attempting
to estimate this fundamental quantity are discussed at length in the classic work by Polyá and
Szegö \cite{polya:szego:1951}. It is well known that the accurate numerical estimation of capacity is quite difficult to
achieve, except few exceptional sets, and that even numerical estimation based on finite
element and other methods can be very challenging. In recent years, however, a powerful and
fast algorithm has been developed that allows for the calculation of ordinary capacity $\Cap_2$ (when $d =3$) to
unprecedented accuracy.  The aim here is to develop an extension of the current algorithm to
describe $\Capalpha$  for the purpose of providing a computational tool for addressing diverse shape
discrimination problems. We emphasize that the method of solution developed by
\cite{kyprianou:osojnik:shardlow:2018} should be valuable in solving for functionals of $K $associated with interior boundary value
problems such the “torsional rigidity” and the related problem of calculating velocity field in pipes
having a general cross-sectional shape as in Hunt et al. \cite{hunt:douglas:bernal:1995}. The methods are thus complementary.

We term our methodology for accelerating the path averaging process for the exterior
Dirichlet problem the Walk-In-Out-of-Balls (WIOB) method. The name of this method derives
taking the starting distribution of launch sites for our paths to be located within a spherical ball
rather from the boundary of such a ball as in the former WOS computations of 2-capacity in the
case of Brownian paths. This distribution of launch sites in our path averaging simulations
derives from the equilibrium charge density of the stable process hitting a solid sphere when
launched from the exterior of the ball. For Brownian paths, the launch points are localized to
the sphere surface because the charge density is localized to the spherical surface, a fact that
greatly simplifies the calculation in this case. This is why the simple walk-on-sphere method can
be readily applied in the Brownian motion for the exterior problem when the launch sphere
concept is utilized to embed the set $K$. We further emphasize that the exterior problem has
other features that make it rather distinct from its interior analog. Capacity is fundamentally
related to probability of the path process to escape to infinity, and thus depends strongly on
the spatial dimensionality in addition to the “shape” and “size” of $K$. Paths initiating from the
interior of (compact) $K$ never escape to infinity so that the transience of the generalized random walk
process (existence of a positive probablity of ultimate escape to infinity) is not a consideration in
interior calculations. In contrast, capacity is only a positive finite quantity when the escape
probability is finite and positive and this limits the values of $\alpha$ and $d$ for which $\Capalpha$  can be
meaningfully defined.

The novelty of our approach is not limited to the development of the WIOB method, which
might be useful in other fractional diffusion exterior boundary value problems. The numerical
calculations of the energy integral involve U-statistics. However, standard U-statistic theory
does not work here because the summands are heavy tailed with infinite variance.
Using standard $U$-statistics theory sometimes gives unreliable
estimates of the energy integral. Indeed, it sometimes gives estimates of variance that are
negative. We also want to be able to detect when the energy integral is infinite, but the
finite sum approximation in
section 3.2 to equation (\ref{eq:energy.integral})  is always finite. We provide a novel approach for solving these numerical
problems that we expect to have general utility in the solution of exterior fractional boundary
value problems. As another point related to the novelty of the present work, we consider a
range of representative numerical computations in dimensions: $d = 2, 3, 4$ and 5. We are not
aware of any numerical estimations of even ordinary capacity in any dimension other than $d =2$ or $3$.
We also explore $\alpha$-capacity when $\alpha < 2$ for hollow objects.
Ordinary capacity cannot probe the interior of hollow sets; $\alpha$-capacity can. We also discuss how $\alpha$-capacity can
be used to estimate the fractal (Hausdorff) dimension of sets and illustrate this type of
calculation in the simple case of a disc shaped region embedded in $d = 3$. We expect that the
capacity to probe information in the interior of sets and to estimate fractal dimension and
other geometrical properties of sets should be useful in numerous future practical applications
regarding shape discrimination and quantification.
% -------------------------------------------------------------------------

We review the analytic definition of capacity given by Riesz \cite{riesz:1938a}.  % pg 131-133
For $\alpha > 0$ and a Borel probability measure $\mu$ on $\reals^d$, the
$\alpha-$energy integral is defined by,
\begin{equation}\label{eq:energy.integral}
I_\alpha(\mu) = \int_{\reals^d}  \int_{\reals^d} \frac{1}{|\boldx-\boldy|^{d-\alpha}} \mu(d\boldx)\mu(d\boldy).
\end{equation}
The kernel of this energy functional derives from the potential associated with L\'evy random walks.
For a compact Borel set $K \subset \reals^d$, the $\alpha$-capacity or Riesz $\alpha$-capacity of $K$ is defined by the functional:
$$\Capalpha(K) = \left( \inf_\mu I_\alpha(\mu) \right)^{-1},$$
where the infimum is over all probability measures $\mu$ with support$(\mu) \subset K$.
Landkof \cite{landkof:1972} shows that there is a unique probability measure $\mu_K$ that minimizes the energy integral, which is
called the $\alpha$-equilibrium measure of $K$.

% -------------------------------------------------------------------------
The classical capacity corresponds to the case in which $\alpha=2$,
and has been variously termed the electrostatic capacity or Coulombic capacity or the Newtonian capacity.
We note that the case $\alpha=d=2$, $\Cap_2(\cdot)$ defined above is finite and it
is not  equal the logarithmic capacity.
We will not discuss the logarithmic capacity in the present work,
see Riesz \cite{riesz:1938a} or Landkof \cite{landkof:1972} for that case.
Also, some authors introduce a multiplicative constant in (\ref{eq:energy.integral}),
which can lead to confusion about the values of $\alpha-$capacity.
In the present work, the  definition of 2-capacity follows the
scaling  used in Riesz \cite{riesz:1938a} and
in the ZENO program of Mansfield and Douglas \cite{mansfield:douglas:2008}
and Juba et al. \cite{juba:audus:mascagni:douglas:keyrouz:2017}.  In particular,
the 2-capacity of a ball of radius $r$ in $\reals^d$ is $r^{d-2}$ for any $d \ge 3$.
We note that unfortunately there is another mathematical quantity that is designated
“$p$-capacity” in the mathematical literature, see Kruglikov \cite{kruglikov:1987}, motivating the
more specific terminology ``Riesz $\alpha$-capacity'' used here.

 Only a few sets have known capacity even when $\alpha=2$ and $d=3$.
 These cases are listed on pages 146-147 of Schiffer and Szego
 \cite{schiffer:szego:1949} and pages 165-167 of Landkof \cite{landkof:1972}.
 Polya and Szego \cite{polya:szego:1951} assert that it is unlikely that there will ever be an analytic method of computing capacity for more general sets.  The work of
Mansfield et al. \cite{mansfield:douglas:garboczi:2001} and Mansfield and Douglas \cite{mansfield:douglas:2008}
developed probabilistic methods based on the paths of Brownian motion for estimating the 2-capacity of
complex shapes in three dimensions $(d=3)$ to understand some of their physical properties.
The resulting algorithm, called ZENO, is based on simulating
Brownian motion paths. The program enables precise calculation of the
2-capacity for complicated objects in dimension $d=3$, e.g. polymers and proteins composed of a large number of atoms.
This general topic is discussed in Vargas-Lara et al. \cite{vargaslara:mansfield:douglas:2017}.
(The program is called ZENO in honor of the famous paradox of Zeno regarding the necessity of taking an infinite number of discrete steps of increasingly small size to arrive at a target endpoint. The ZENO program resolves a ``paradox'' of a similar nature by introducing a cut-off threshold distance to determine when and where the probing random walk path endpoint first ``arrives'' at the set K.)

The purpose of
this paper is to generalize these methods based on Brownian motion to compute the $\alpha-$capacity $\Capalpha$ of complex sets in dimension
$d \ge 2$ with $0 < \alpha \le 2$, using  $\alpha-$stable processes.
% The $\alpha$ range in which $\Capalpha$ is defined depends on $d$, as we discuss below.
The method is based on  hitting probabilities and locations for balls for such a process.

In Section 2, we describe how $\alpha$-stable motion can be used to approximate the
equilibrium measure for the cases $0 < \alpha < 2$.
It is our hope that our novel path sampling method will be useful for other exterior extensions
of classical boundary value problems involving the replacement of the Laplacian operator by a
fractional Laplacian operator $(-\Delta) ^{\alpha/2}$ and the corresponding Levy path processes
of fractal (Hausdorff) dimension $\alpha$ "generated" by fractional Laplacian operators of index $\alpha$.

The third section discusses computing $\alpha-$capacity from
the hitting locations, including numerical considerations and confidence interval estimates
from $U$-statistics.  Section 4 gives some numerical examples of $\alpha-$capacity.
We end with a conclusion and an Appendix that collects technical results.

% -------------------------------------------------------------------------
\section{Simulating hitting locations for stable paths}
% -------------------------------------------------------------------------

In this section, we discus how to simulate hitting locations starting from infinity for a general Borel set $K$, based on
the knowledge of what happens with a ball.  The reason this is of interest is because
this hitting distribution is equal to the capacity equilibrium distribution, see Theorem~2 in Port \cite{port:1967}.
In the next section we will describe how to compute $\alpha$-capacity from that hitting information collected over
many simulated paths.

\begin{figure}
  \centering
  \includegraphics[width=\textwidth]{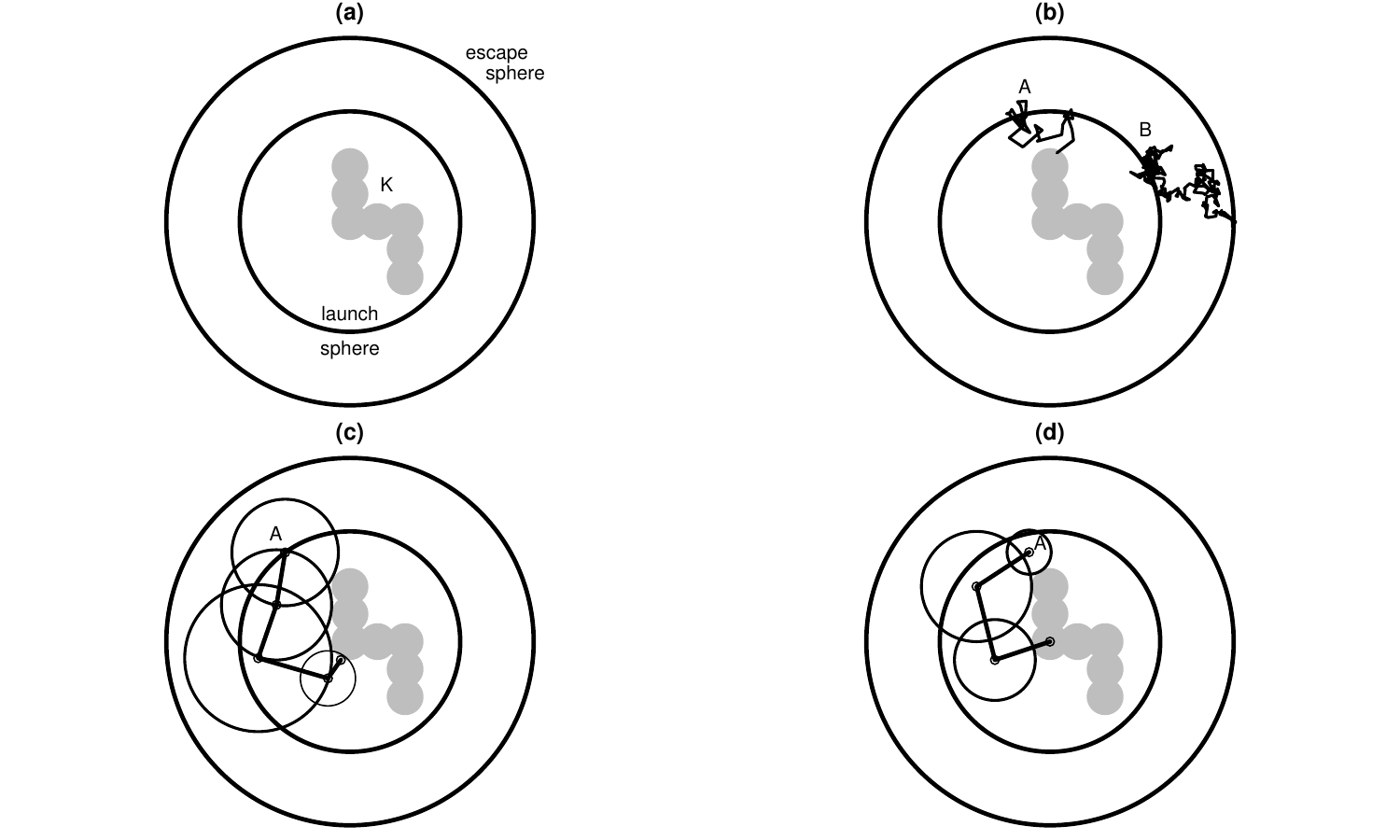}
  \caption{ Types of random walk models in two dimensions.
  (a) This plot shows the setup with an object $K$, a launch sphere (inner circle) containing $K$, and escape sphere (outer circle).
  (b) Two paths for a simple random walk when $\alpha=2$.  One starts at point A on the launch sphere and after multiple steps, hits (comes within $\epsilon$ of) $K$.  The other starts at point B and
  wanders for a while before crossing the escape sphere.  In this case, we decide whether the path escapes to infinity or whether it reenters the launch sphere as in step 3 of the algorithm.
  (c) A Walk-On-Spheres path for Brownian motion starting at point A on the launch sphere.  Successive points are chosen randomly from the sphere just touching the set $K$. In this case,
  the path takes 4 steps and ends up within $\epsilon$ of $K$ and stops.
  % \rred{(This is the walk-on-spheres method, where the   path is NOT as erratic as Br. motion.)}
  (d) A Walk-In-and-Out-of-Balls path for an $\alpha$-stable walk starting at point A inside the launch sphere.    In general, each step will go outside the sphere
  just touching $K$, and eventually the walk will either end up inside $K$ or within $\epsilon$ of $K$ or  escaping to infinity.
  }\label{fig:types.of.walks}
\end{figure}

Let $\ball$ be the (solid) unit ball in $\reals^d$ and then $\boldx + r \ball$ is the ball with center $\boldx$
and radius $r$. Throughout we assume that there are two radii $0 < \Rlaunch < \Rescape$.
See Figure~\ref{fig:types.of.walks}(a) for the $d=2$ case.
It is assumed that the object $K$ is contained in the launch ball $\Rlaunch \ball$.
The larger sphere of radius $\Rescape$ will be called the escape sphere; it is used in deciding whether a random
walk escapes to infinity without hitting $K$.

For both Brownian and stable paths  a probabilistic description of where the walk
exits or enters a ball is known.
Since Brownian motion has continuous sample paths, the enter or exit location will be on the surface of
the ball with probability 1.  However, when $0 < \alpha < 2$, the paths are discontinuous and the hitting
location will be strictly inside or outside the ball with probability 1;
see Blumenthal et al. \cite{blumenthal:getoor:ray:1961}.
If we start a path at a finite location, the Appendix shows
the distribution of hitting locations for a ball.
%  \textcolor{red}{Is this useful?} Scaling a set will change the relative values of $\alpha_1$-capacity
%  and $\alpha_2$-capacity.    Assume $0 < \alpha_1 < \alpha_2 \le 2$ and $\Cap_{\alpha_1}(K) > 0$.  Then $\Cap_{\alpha_2}(K) > 0$ and for any $r > 0$,
%  $$\Cap_{\alpha_1}(r K) / \Cap_{\alpha_2}(rK) = r^{\alpha_1 - \alpha_2} \Cap_{\alpha_1}(K) / \Cap_{\alpha_2}(K) .$$
%  For small enough $r$, the ratio is greater than 1, for large enough $r$ the ratio is less than 1.

To exactly simulate the hitting from infinity location, we need to choose an appropriate finite starting distribution.
The correct starting location is based on where a walk starting at infinity hits the launch ball.
For the Brownian case, this means the starting location is
on the launch sphere and that distribution is uniform; for $0 < \alpha < 2$ cases, the starting location
is distributed across the launch ball according to the
equilibrium distribution for the launch ball, e.g., equation (\ref{eq:equilibrium.dist}) in the Appendix.

% -------------------------------------------------------------------------
\subsection{Stable random walks}\label{sec:simple.rw}
% -------------------------------------------------------------------------

Mansfield et al. \cite{mansfield:douglas:garboczi:2001} introduced an algorithm for computing the 2-capacity in the special case
where $\alpha=2$ and $d=3$ based on hitting the set $K$ with simple random walk trajectories.  As noted above, these
paths start uniformly on a launch sphere that contains $K$.  We next describe a natural extension  of this
algorithm to general $\alpha$ and dimensionality  $d$.  Throughout we assume $\alpha < d$ to guarantee
that the L\'{e}vy flights are transient (see Appendix for definition).
%We now describe how to generalize this to $\alpha$-capacity.
The method requires two basic elements: a way to simulate isotropic stable random vectors and to be able to
compute the probability of hitting a ball given that we start at an exterior point.
The Appendix describes how to do both of these for $\alpha$-stable motion in $\reals^d$, $0 < \alpha \le 2$.

Let $\gamma > 0$  be a scale variable for the step size, i.e.
each step $\boldY$ of the random walk will have characteristic function
\begin{equation}
E\exp( i \boldu \cdot \boldY ) = \exp( -\gamma^\alpha |\boldu|^\alpha ). \label{eq:isotropic.stable.step}
\end{equation}
This isotropic stable process is described in Chapter 3 of Sato \cite{k.sato:1999}.  Let $\epsilon \ge 0$ be a hitting tolerance, i.e. we will consider that the walk
has hit $K$ if the distance between a point and $K$ is less than or equal $\epsilon$.
Typically $\epsilon  \ll \gamma \ll \Rlaunch$; the first inequality implies that the $\epsilon$ ``thickening'' of $K$
is small compared to the typical step size and the second inequality implies that the typical step size is small
compared to the size of the object.  We generate a path according to the following
algorithm, see Figure~\ref{fig:types.of.walks}(b).
\begin{enumerate}
  \item Pick a starting point $\boldx$.   If $\alpha=2$, this means pick $\boldx$ uniformly on the launch sphere; if $\alpha < 2$,
  this means pick $\boldx$ according to the equilibrium distribution on the launch ball using equation (\ref{eq:equilibrium.dist}) below.
  \item Generate a step  $\boldy$ with distribution given by equation (\ref{eq:isotropic.stable.step}) and compute a new location $\boldx \leftarrow \boldx + \boldy$.
  \item If $|\boldx| > \Rescape$, compute the probability $p$ of a walk hitting the launch ball starting at $\boldx$.
  (Recall that the process is transient.)  With probability $1-p$,
  declare that this walk will escape to infinity without ever hitting the launch ball (and therefore not hit $K$) and we stop this path.  Otherwise, compute a new location $\boldx$ by
  simulating a hitting location in the launch ball by a walk starting at the current location as described in the Appendix.
  \item If $\boldx$ is within $\epsilon$ of $K$, declare the current location $\boldx$ to be a hit.  Otherwise, go to step 2.
\end{enumerate}

We note that since a stable process generally jumps into the interior of a set, we can set $\epsilon=0$ for the
kinds of sets considered here.  For thin sets, say fractals, we may want to use an $\epsilon > 0$ to effectively
``thicken'' $K$, but the influence of this on the final results should be carefully examined.

This process is repeated $M$ times and a count of number of hits of $K$ and their locations are saved.

% -------------------------------------------------------------------------
\subsection{Walk-On-Spheres method for Brownian motion}
% -------------------------------------------------------------------------

When $\alpha=2$,  Muller \cite{muller:1956} introduced a faster method to simulate from the hitting distribution of a general set $K$.  The continuity of Brownian paths
guarantees that a path can only leave a ball by crossing its boundary, a sphere.
Furthermore, the rotational symmetry means that the distribution
of exit points is uniform on the sphere.  So rather than simulate many steps, this method
chooses random exit points on a sequence of (random) spheres, see Figure~\ref{fig:types.of.walks}(c). This is called
the Walk-On-Spheres method, abbreviated WOS.  Here are the steps in the algorithm
to generate a single path.

\begin{enumerate}
  \item Pick a point $\boldx$ uniformly from the launch sphere.
  \item Compute $r \leftarrow |\boldx-K| = \inf_{\boldk \in K} |\boldx-\boldk|=$ distance between $\boldx$ and $K$.
  \item Pick a new point $\boldy$ uniformly on the sphere of radius $r$ centered on $\boldx$
  \item If $|\boldy-K| \le \epsilon$, return $\boldy$ as the hitting location.
  \item Otherwise, set $\boldx \leftarrow \boldy$.
  \item If $|\boldx| > \Rescape$, compute the probability $p$ of a walk starting at $\boldx$ hitting the launch ball.  With probability $1-p$,
  declare that this walk will escape to infinity without ever hitting the launch ball (and therefore not hit $K$).
    Otherwise, go to step 2.
\end{enumerate}

% -------------------------------------------------------------------------
\subsection{Walk-In-and-Out-of-Balls for stable path processes}\label{sec:WIOB}

When $0 < \alpha < 2$, the Walk-On-Spheres method cannot be used because the stable process has jumps and will exit a ball not on the boundary,
but on some point exterior to the ball.  Blumenthal et al. \cite{blumenthal:getoor:ray:1961} give expressions for
the distribution of these exit point for $\alpha$-stable processes with $0 < \alpha < 2$.
They also give the probability of hitting a ball when starting from some exterior point and what the hitting location is
on the interior of the ball.  These expressions are given in the Appendix.

The approach below extends the work of Kyprianou et al. \cite{kyprianou:osojnik:shardlow:2018}.
That work considers the interior problem: a stable process starts at a point $\boldx \in K$, and the paths are followed
until they exit $K$.  In contrast, we are concerned
with the exterior problem:  in which we start at a point $\boldx \not \in K$, and propogate the path until it hits $K$.
There are several key differences between \cite{kyprianou:osojnik:shardlow:2018} and the method proposed here; we describe them below.

Our approach generates a sequence of points that are outside a sequence of balls, with a check to avoid
drifting too far from the object that sometimes returns us to the launch ball $\Rlaunch \ball$.
We call this method the Walk-In-and-Out-of-Balls method, abbreviated WIOB.
The steps of the method are given below, see Figure~\ref{fig:types.of.walks}(d) to visualize the steps.

\begin{enumerate}
  \item Pick a point $\boldx$ on the launch ball according to the equilibrium distribution in equation (\ref{eq:equilibrium.dist}).
  \item Compute the distance $r=|\boldx-K|$ = distance between $\boldx$ and $K$.
  \item Simulate an exit location $\boldy$ from the ball of radius $r$ centered on $\boldx$.
  \item If $|\boldy-K| \le \epsilon$, return $\boldy$ as the hitting location.
  \item Otherwise, set $\boldx \leftarrow \boldy$.
  \item If $|\boldx| > \Rescape$, compute the probability $p$ of a walk starting at $\boldx$ hits the launch ball. With probability $1-p$,
  declare that this walk will escape to infinity without ever hitting the launch ball (and therefore not hit $K$)
  and stop this iteration. Otherwise, simulate reentering the launch ball at a new point $\boldx$ and go to step 2.
\end{enumerate}

There are two  advantages of this method.  First, like the WOS method for Brownian motion, it is generally more efficient.
Instead of many small steps, the focus on leaving a ball means each WIOB step is larger and faster than many small steps.
Second, WIOB allows a miss of the object $K$ that could occur with a jump that the WOS method does not allow.  The chance of this
happening is small, because small jumps are more likely, but the probability of a large step increases as $\alpha$ decreases.

We now describe differences between \cite{kyprianou:osojnik:shardlow:2018} and WIOB.  Both methods
rely on the key results in \cite{blumenthal:getoor:ray:1961}, but have different objectives.
\cite{kyprianou:osojnik:shardlow:2018} is interested in simulating paths to compute a solution of a fractional diffusion equation and do not
consider capacity, which is the main focus of this paper.
They fix a starting location $\boldx$, whereas we start at a random location,
which we specified above, chosen to exactly simulate hitting from infinity in step 1.
They do not have to deal with transience and the possibility that
the process never hits $K^C$, whereas we deal with this possibility in step 6. Simulating
the return to the launch ball uses the density $f_{BGR}$ from the Appendix.
The straightforward way of doing this with simple rejection is increasingly less efficient as the
point becomes closer and closer to the launch ball.    Devroye and Nolan
\cite{devroye:nolan:2022} have developed algorithms for exactly simulating locations for entering or leaving a ball
 that are uniformly bounded for all starting locations $\boldx$.
We note that these methods work for all  $0 < \alpha \le 2$  and any dimension $d \ge 2$, enabling calculating
standard capacity and Riesz capacity.  This enables new applications in dimension not equal to 3.

% -------------------------------------------------------------------------
\section{Estimating $\Capalpha$ from hitting locations}
% -------------------------------------------------------------------------

Throughout this section $K \subset \reals^d$ is a fixed compact set.
We will use the results of the previous section to compute an estimate $\widehat{\Capalpha}(K)$ of  $\Capalpha(K)$
and give confidence intervals for this estimate.

% -------------------------------------------------------------------------
\subsection{Electrostatic capacity from Brownian paths}

For electrostatic capacity, e.g. $\alpha=2$ and $d=3$, then for any $|\boldx| > 1$,

\begin{equation}\label{eq:hit.ball.alpha.2}
P(\boldX_t \text{ hits the unit ball} \, \ball \, | \, \boldX_0=\boldx ) = \frac{1}{|\boldx|}.
\end{equation}
Based on this result, it can be shown that if $K$ is any compact set,
\begin{equation}\label{eq:hit.prob.alpha.eq.2}
\Cap_2(K) =  \Rlaunch \, P(\boldX_t \mbox{ hits } K | \boldX_0 \text{ is uniform on } \{ |\boldx|=\Rlaunch \}  ).
\end{equation}
% This follows from Theorem~8.27 and Theorem~8.33 of \cite{morters:peres:2010}.

The sample analog is straightforward: run a large number of random walks and
count the number of hits of $K$ and divide by the number of paths simulated.
This is the essential idea underlying the ZENO program described in %\cite{juba:audus:mascagni:douglas:keyrouz:2017}.
Mansfield et al. \cite{mansfield:douglas:garboczi:2001} and Mansfield and Douglas \cite{mansfield:douglas:2008}.

% -------------------------------------------------------------------------
\subsection{Riesz capacity $\Capalpha$ from $\alpha$-stable sample paths}

When $0 < \alpha < 2$, the situation is more complicated.
For a ball, there is an expression for the hitting probability that is the analog  of equation (\ref{eq:hit.ball.alpha.2}),
see equation (\ref{eq:hit.ball.prob.gen.alpha}) in the Appendix below.
However, there is no known relationship between hitting probability and $\alpha$-capacity for a general set $K$.

The proposed solution is to run $n$ simulations of the WIOB method to obtain hitting locations $\boldx_1,\ldots,\boldx_n$ in the set $K$.
Using these hitting locations gives an empirical distribution $\widehat{\mu}_K$ that is an
approximation to the equilibrium measure of the set.  The straightforward approach is to evaluate the energy integral
(\ref{eq:energy.integral}) with a sum, i.e.
$$I_\alpha(\widehat{\mu}_K) = \frac{1}{\binom{n}{2}} \mathop{\sum  \sum}_{i < j} \| \boldx_i - \boldx_j \| ^{\alpha-d},$$
where $\binom{n}{2}$ is the number of distinct non-diagonal pairs in $\{ \boldx_i - \boldx_j \}$.

Unfortunately, there are two problems with this seemingly straightforward approach.   The first problem is
that this sum is a numerically unreliable way to approximate the energy integral because the integrand  in (\ref{eq:energy.integral}) has a singularity on the
diagonal $\boldx=\boldy$.   With a sample, if $\boldx_i \approx \boldx_j$, one or more terms in the sum above can dominate the sum, leading to a poor estimate of $I_\alpha(\mu)$.  The second problem is that the terms in this sum are dependent, so it is not straightforward to get a confidence interval
for such an estimate.  Below, we describe a way to split the sample into two groups and use different estimators form
the groups in a way that gives
less volatile estimates of capacity and valid confidence intervals.

An equivalent way of viewing this is to define the random variable $W = \| \boldX - \boldY \|^{\alpha-d}$, where $\boldX$ and $\boldY$ are independent copies with distribution $\widehat{\mu}_K$.
Then $I_\alpha(\widehat{\mu}_K) = E W$, the expectation of $W$.
% Points on $K$ are within distance  diameter$(K)$ of each other, so $0 < \mbox{diameter}(K)^{\alpha-d}  \le W < \infty$.
Points that are close to
each other on $K$ correspond to large values of $W$.  In probabilistic terms, $W$ is a heavy-tailed variable and therefore estimating the mean can be unreliable.
And contrary to most statistical estimates, the larger the sample, the more likely it is that we will see very large values of $W$ and get a less reliable estimate of $EW$.
Furthermore, in cases where the $\Capalpha(K)=0$, $EW=\infty$, but the sample mean will never lead to a value of $\infty$.

We propose a method to approximate $EW$ that detects the $EW=\infty$ case, gives reliable estimates of
$I_\alpha(\widehat{\mu}_K)$ when it is finite, and
in this later case, gives a large sample confidence interval for $I_\alpha(\mu)$, and thus a confidence interval for $\Capalpha(K)$.
Start with a sample of $n$ hitting locations $\boldx_1,\ldots,\boldx_n$ and define $w_{ij} = \| \boldx_i - \boldx_j \|^{\alpha-d}$, $i \ne j$.
Pick a threshold $\tau > 0$ for determining which of these values is large.  A convenient choice is to define $\tau$ to be an upper quantile of $W$,
e.g. $P(W \le \tau) = p_\tau$ for some value $p_\tau \approx 1$.
Split the integral for $E W$ into two pieces:
$$EW = E \min(W,\tau) + E \max(0,W-\tau) = I_1 + I_2.$$
We will estimate $I_1$ using $n_1$ of the $n$ samples and $I_2$ by a tail estimator based on the remaining $n_2=n-n_1$ values.

For $I_1$, define $W^* = \min(W,\tau)$.  This is a bounded random variable, so necessarily has a
finite mean and variance.  Then, define the sample analogue $w^*_{ij} = \min( w_{i,j}, \tau )$ for $1 \le i \ne j \le n_1$
and the following terms:
\begin{eqnarray*}
  \overline{x}_{j} & = & \frac{1}{n_1-1} \sum_{i \ne j}^{n_1}  w^*_{ij} \\
  \widehat{I}_1     & = & \frac{1}{n_1} \sum_{j=1}^{n_1} \overline{x}_{j}
                             = \frac{1}{n_1(n_1-1)} \sum_{j=1}^{n_1} \sum_{i \ne j}^{n_1}  w^*_{ij} \\
  v_1 & = & \mathrm{Var}( \overline{x}_{j} ) = \frac{1}{n_1-1}
                    \sum_{j=1}^{n_1}   \left( \overline{x}_{j}^2 - \widehat{I}_1 \right)^2
\end{eqnarray*}
Then $\widehat{I}_1$ is a U-estimator of $I_1$.  The variance term $v_1$ is used below to get a confidence interval for $I_1$.

For $I_2$, we assume that the upper tail of $W$ has a power decay:
\begin{equation} \label{eq:tail.condition}
P(W > w) \approx c w^{-\nu} \textrm{  for large } w.
\end{equation}
To estimate $\nu$, select values $z_1,\ldots,z_{n_3}$ from $w_{ij}$, $n_1 < i,j \le n$ where  (a) $w_{ij} > \tau$, (b) there is at most
one value selected from each row of the $w_{ij}$ matrix, and (c)  there is at most
one value selected from each column of the $w_{ij}$ matrix. Since we are only choosing the $z_i$ from
$i,j > n_1$, the $z_i$ are independent of the $w_{ij}$ used in estimating $I_1$, the values $z_i$ are independent of $I_1$
and because of conditions (b) and (c) above, the $z_i$ values are independent of each other.
Hence, the $z_i$ values are an i.i.d. sample from the upper tail of $W$.  We use the Hill estimator for the tail decay $\nu$:
$$\widehat{\nu} = \left( \frac{1}{n_3} \sum_{i=1}^{n_3} \log z_i - \log \tau \right)^{-1}.$$
In the simulations we have run, we've observed $\widehat{\nu} \le 2$ when $d \ge 3$ or ($d=2$ and $\alpha \le 1$), in which case Var$(W) = \infty$.
Assume that the tail approximation (\ref{eq:tail.condition}) is exact for $W > \tau$.  Then $W$ has an upper tail density
$c \nu w^{-\nu-1}$ and necessarily $c= (1-p_\tau) \tau^\nu$.
Then the value of $I_2$ is $\int_{\tau}^\infty (w-\tau) c \nu  w^{-\nu-1} = (1-p_\tau) \tau/(\nu-1)$ and the  estimate of $I_2$ is
$$\widehat{I}_2 = \begin{cases} (1-p_\tau) \tau/(\widehat{\nu}-1) & \widehat{\nu} > 1 \\
                        +\infty & \widehat{\nu} \le 1.
         \end{cases} $$

The energy integral estimate is $\widehat{I} = \widehat{I}_1 + \widehat{I}_2$,
and the $\alpha$-capacity estimate is $\widehat{\Capalpha}(K) = 1/\widehat{I}$.
Note that if $\nu \le 1$, the sum $I_1 + I_2 = \infty$ and $\widehat{\Capalpha}(K) = 0$.

% -----------------------------------------------------------------------
\subsection{Confidence interval for $\alpha$-capacity estimates}\label{sec:conf.intervals}

The ZENO program has a method of deriving large sample confidence intervals when $\alpha=2$ and $d=3$.
Large sample confidence intervals based on the normal distribution are described next for arbitrary $d \ge 2$ and $0 < \alpha \le 2$.

%The $\alpha=2$ case is straightforward:  (\ref{eq:hit.prob.alpha.eq.2}) shows that a confidence interval can be based on a scaled binomial estimator $\widehat{p}$ for
%$p=P(\boldX_t$ hits $K)$. Standard statistical theory shows that for large $n$, $\widehat{p}$ is approximately N$(p,p (1-p)/n)$, and thus the estimator based on
%(\ref{eq:hit.prob.alpha.eq.2}) is approximately N$(p \Rlaunch,p (1-p) \Rlaunch^2 /n)$.  Plug in the sample value of $\widehat{p}$ and known value of $\Rlaunch$ to get a
%large $n$ confidence interval for $\Capalpha(K)$:  $\widehat{p} \Rlaunch \pm  z_{\alpha/2} \sqrt{ \widehat{p} (1-\widehat{p}) \Rlaunch^2/n}$.
%%\rred{Check that this is what ZENO does.}

%The rest of this section will focus on the case $\alpha < 2$.
We assume $\nu > 1$, so $I < \infty$.
First, the variance of $\widehat{I}$ will be estimated.
By the way the sample is split above, the two estimators $\widehat{I}_1$ and $\widehat{I}_2$ are independent,
so Var$(\widehat{I})$ =  Var$(\widehat{I}_1)$ +  Var$(\widehat{I}_2)$.  The standard theory of U-statistics,
e.g. Chapter 12 of van der Vaart \cite{vandervaart:1998}, gives an estimator of the variance of $I_1$ that is sometimes negative.
(This possibility is mentioned on page 419 of Schucany and Bankson \cite{schucany:bankson:1989}; surprisingly,
this can happen with large samples,
e.g. $n_1=10000$.)  To avoid that problem, we use the modified variance estimator of
Sen \cite{sen:1960}: namely $ \sigma_1^2 = \text{Var}(\widehat{I}_1) = 4 v_1/n_1.$

With appropriate regularity conditions, the Hill estimator $\widehat{\nu}$ based on the $n_3$ largest values is approximately
N$(\nu,\nu^2/n_3)$, see page 304 of Resnick \cite{resnick:2007}.
%If the tail condition (\ref{eq:tail.condition}) is exact for $w > \tau$, then
%we necessarily have $c=\tau^\nu$ and tail integral
%$E(W 1_{W > \tau}) = \int_\tau^\infty w \tau^\nu /w^\nu dw =  \tau \nu/(\nu-1)$.  This shows why $\widehat{I}_2$ has the given form.
Using the delta method on $g(\nu)= (1-p_\tau) \tau /(\nu-1)$ shows that $\widehat{I}_2$ is approximately N$(I_2,\sigma_2^2)$, with $\sigma_2^2=g^\prime(\nu)^2/n_3$.
Thus $\widehat{I} = \widehat{I}_1 + \widehat{I}_2$ is approximately
N$(I,\sigma_I^2)$, where $\sigma_I^2=\sigma_1^2+\sigma_2^2$.  Finally, using the delta method again, the
 estimate of $\alpha-$capacity $\widehat{\Capalpha} = 1/\widehat{I}$ is approximately N$(\Capalpha,\sigma_{\Capalpha}^2)$, where
$\sigma_{\Capalpha}^2 =\sigma^2_I / I^4$.  Hence, a large sample $(1-\delta)$ confidence interval is given by
$\widehat{I} \pm z_{\delta/2} \sigma_I/\widehat{I}^2$.

% -------------------------------------------------------------------------
\section{Implementation and numerical examples}

The Walk-In-and-Out-of-\linebreak Balls method has been implemented in R, a free open source
programming language, see R Project \cite{RProject}.  It allows general $d \ge 2$ and
some basic types of target objects: unions of balls, unions of cubes, unions
of rectangular solids, and cylinders.
The simulations in this  sections shows some examples of estimated $\alpha$-capacity for
several objects in different dimensions.

% ---------------------------------------------------------------------
\subsection{$\Capalpha$ of a ball in $\reals^d$}

\begin{figure}
  \centering
  \includegraphics[width=\textwidth]{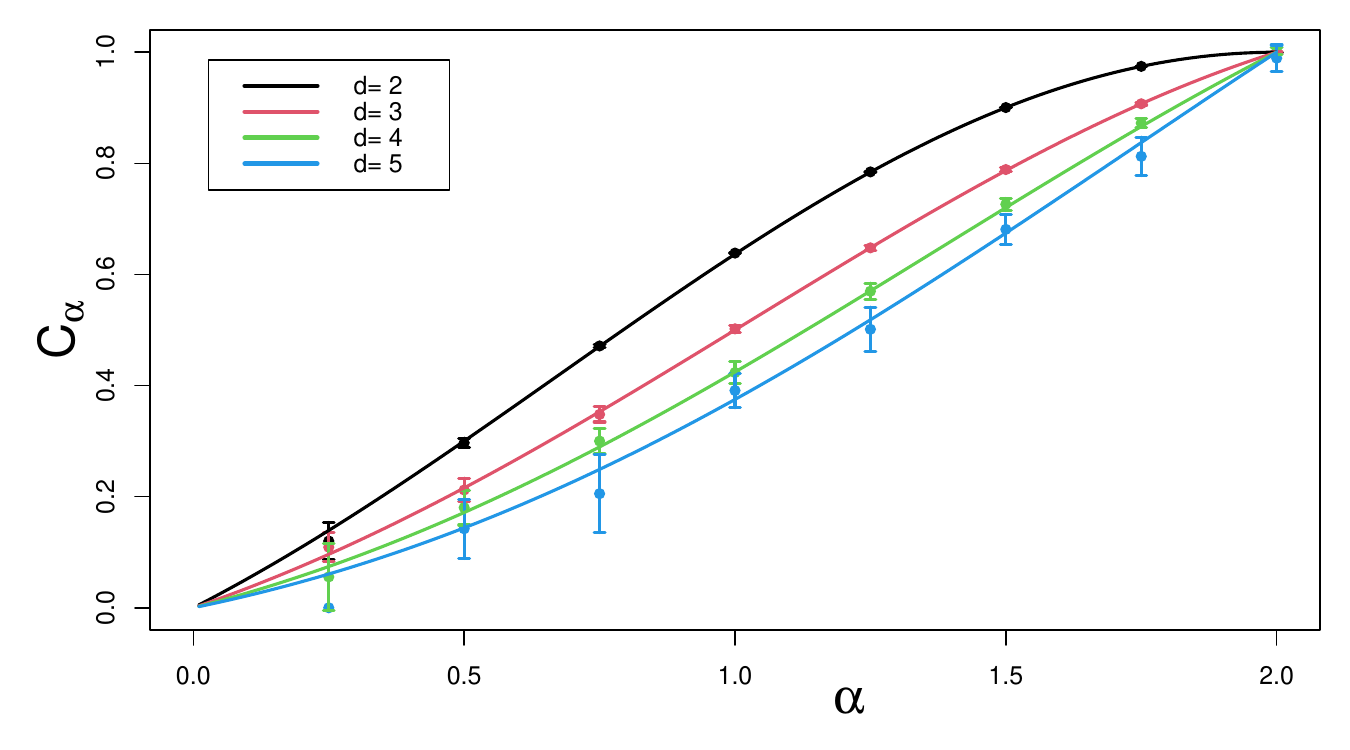}
  \caption{Estimated $\Capalpha$ of the unit ball for  dimensions $d=2,3,4,5$ as a function of $\alpha$.  The circles are estimated
  capacity using hitting locations simulated by a stable random walk, the solid curves are the exact capacity.
  %\rred{The solid curves ar exact analytic values, not estimated!}
   The estimates were based on $n=10000$ hits of the ball,
  with $n_1=5,000$, $\epsilon=10^{-6}$, $\Rlaunch=2$, $\Rescape=4$, and $p_{\tau}=0.995$.  The error bars on the plot give 95\% confidence intervals; they are small for $d=2$ or $3$, more visible when $d=4$ or $5$.  Each point required approximately 1 minute of time on a 2.3 GHz CPU.}\label{fig:ball_capacity}
\end{figure}

The only case where an exact formula is known for the $\alpha$-capacity of a $d$-dimensional set is for a ball;
 it is given in equation (\ref{eq:ball_capacity}) below.
 Here we focus on the unit ball in dimension $d=2,3,4,5$.  This first simulation
used the  WIOB method to estimate $\alpha$-capacity and then compares those estimates to the
theoretical values.  Figure~\ref{fig:ball_capacity} compares these numerical estimates to the exact theoretical
values in equation (\ref{eq:ball_capacity}) in the Appendix.  Since the
exact $\alpha$-capacity of a sphere is known, we may gauge the accuracy of the simulation by the deviation between estimates based on simulation
 and analytic, exact values.  For example, when $d=3$, the confidence interval width varies from 0.0948 when $\alpha=0.25$
 to 0.00318 when $\alpha=2$.  Increasing the sample size will improve the accuracy.

% ---------------------------------------------------------------------
\subsection{$\Capalpha$ of solid and hollow cubes}\label{sec:solid.hollow.cubes}

\begin{figure}
  \centering
  \includegraphics[width=\textwidth]{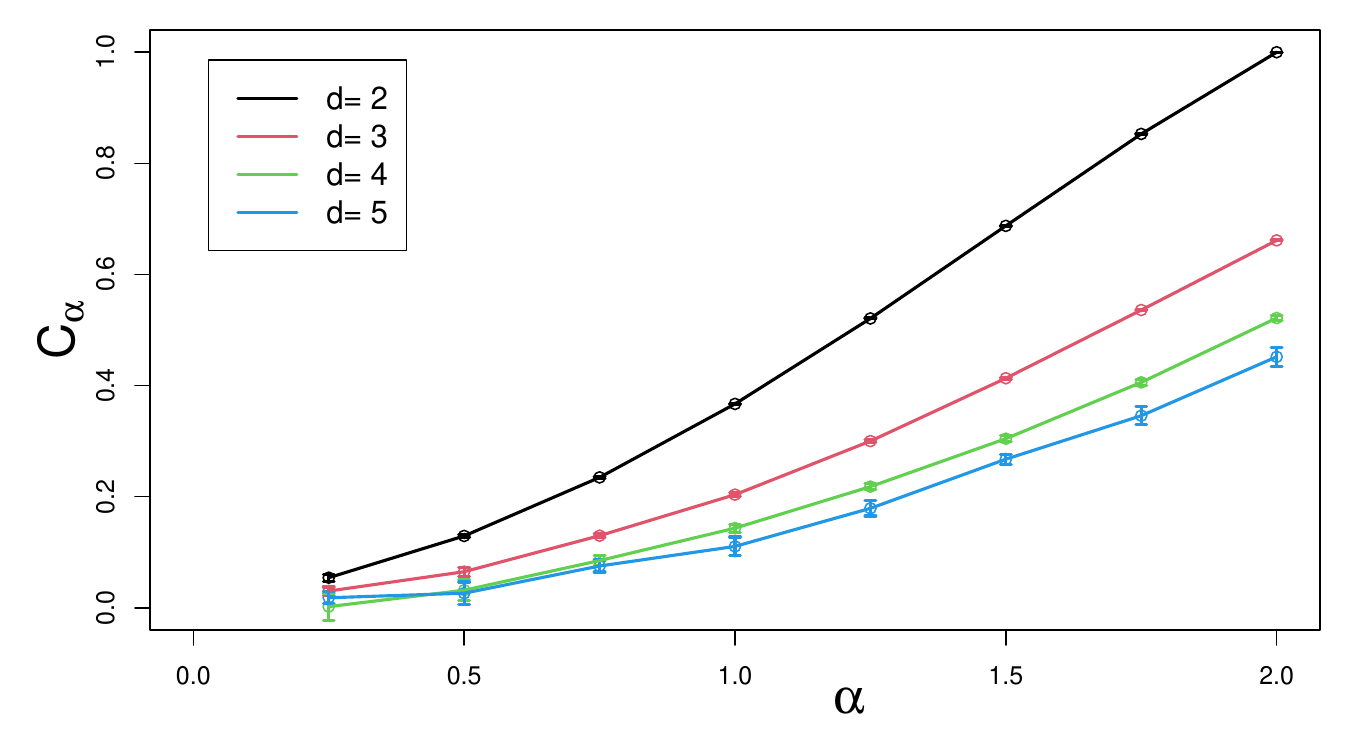}
  \caption{Estimated $\Capalpha$ of a solid cube $[0,1]^d$ for  dimensions $d=2,3,4,5$ as a function of $\alpha$.
  The solid lines connect estimated values.  Parameters are the same as in Figure~\ref{fig:ball_capacity}. }\label{fig:cube_capacity}
\end{figure}

First let $K$ be the solid cube $[0,1]^d$.
Figure~\ref{fig:cube_capacity} shows estimated $\Capalpha$ of a these cubes
in dimensions 2, 3, 4, and 5 as a function of $\alpha$.
In particular, the estimated value of this capacity when $d=3$ and $\alpha=2$ is $0.66134  \pm 9.7 \times 10^{-4}$
using WIOB with $n=10000$.  This compares to previous estimates of
0.663  in Mansfield et al. \cite{mansfield:douglas:garboczi:2001},
$0.660675  \pm  10^{-5}$  using 4.7 billion trajectories in Given et al. \cite{given:hubbard:douglas:1997}
and   $0.6606782 \pm 10^{-7}$  in Hwang et al. \cite{hwang:mascagni:2004}.
The particular virtue of the path integration method for estimating capacity is that it allows for highly
precise estimates in comparison with other methods such as finite element methods when a large number of trajectories is employed.

Separate simulations, not shown here, demonstrate that the method introduced here satisfies the scaling property given in
equation (\ref{eq:translation.n.scaling}) in the Appendix:
 $\Capalpha( r K ) = r^{d-\alpha} \Capalpha( K )$ for $r > 0$.  So these capacity values for
unit cubes give the capacity of any cube.

\begin{figure}
  \centering
  \includegraphics[width=\textwidth]{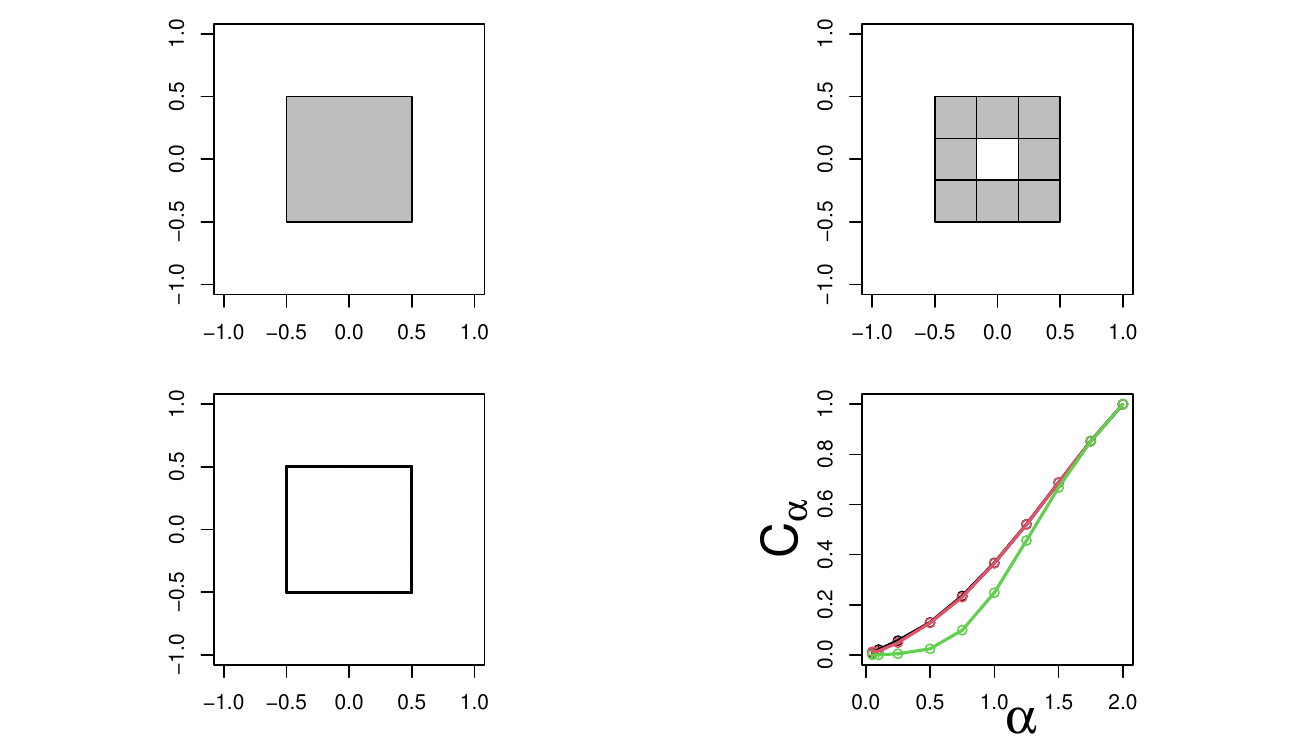}
  \caption{Cubes and their estimated $\alpha$ capacity.  Top left is the centered solid unit cube $[-1/2,1/2]^2$ , top right is a thick walled cube and bottom left
  is a thin walled cube.  The bottom right plot shows $\alpha$-capacity.
  The capacity for the solid cube (black) and for the thick walled hollow cube (red) are virtually indistinguishable,
  but the thin walled
  cube (green) increasingly diverges from the other two as $\alpha$ decreases.}\label{fig:hollowcube_capacity}
\end{figure}

Since non-Gaussian stable processes exhibit discontinuous path displacements, their hitting locations are not restricted to the surface of an object.
This class of erratic path processes can thus penetrate through the boundary before they hit $K$.
So unlike 2-capacity, the $\alpha$-capacity of a hollow object can differ from the $\alpha$-capacity of a solid object
with the same bounding surface.
 To examine this effect in 2-dimensions, we look at a solid and hollow squares.  Figure~\ref{fig:hollowcube_capacity}
shows these three regions: the first (upper left) is a single, solid square $[-1/2,1/2]^2$ (top left), the next is a hollow square with a thick
wall of width 1/3 (top right), and one with a thin wall of width 1/100 (bottom left).  The bottom right plot shows the estimated $\alpha$-capacity of
the three objects for a range of $\alpha$ in (0,2].  The black curve for the solid cube is mostly covered by the thick walled curve in red, showing that it
is difficult to get through a thick barrier when most of the jumps of a stable process are small.  For the thin-walled cube, the green curve does diverge from
the other two curves, and this difference is larger when $\alpha$ is small.  In theory, this allows one to distinguish between solid and hollow objects with
the same surface, something not possible with Brownian motion.

% -------------------------------------------------------------------------
\subsection{$\Capalpha$ of a solid rectangular bar in three dimensions}

\begin{figure}
  \centering
  \includegraphics[width=\textwidth]{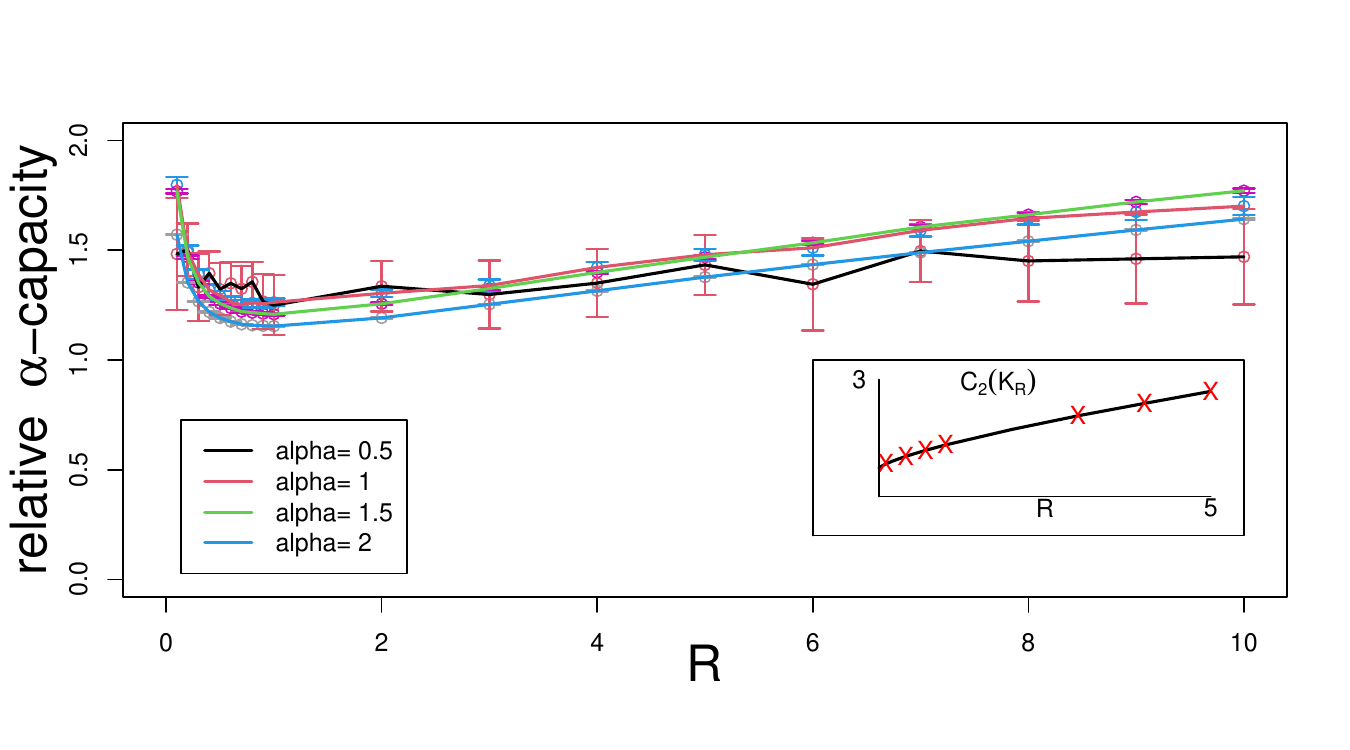}
  \caption{The main plot shows the relative capacity $\Capalpha(K_R)/\Capalpha$(ball with the same volume as $K_R$)  for solid bars in 3 dimensions.
 The inset (lower right)
  compares two estimates of the 2-capacity of $K_R$;  the solid black curve is from the ZENO program while the
  red x's show estimates from the WIOB algorithm.
  The simulations used n=10000 hits of $K_R$, $\epsilon=$ 1.0e-6,  $n_1= n/2=5000$, $p_\tau=0.9999$, and confidence interval level
$\delta=0.95$.   }\label{fig:rod.capacity}
\end{figure}

Here we consider an anisotropic example in $\reals^3$:  $K_R= [-R,R] \times [-1,1]^2$.
This is a solid bar with length $2R$ and  cross section a square of area 4, so it has volume $8R$.
As $R$ varies away from 1 in either direction, the set $K_R$ gets more and more anisotropic.
Since the volume of $K_R$ increases with $R$, we normalize each capacity estimate by dividing it by
the $\alpha$-capacity of a ball with the same volume as $K_R$.
This type of dimensionless ratio, the relative capacity, provides valuable information about
the shape of a region, e.g., Polya and Szego \cite{polya:szego:1951}.
Figure~\ref{fig:rod.capacity} shows the results of
simulating hitting locations and computing $\alpha-$capacity for a range of $\alpha$ and $R$.
The increase in relative capacity near $R=0$ is caused by the fact that $\Capalpha(K_R)$ decreases to 0 more slowly
  than $\Capalpha$(ball with the same volume as $K_R)$. The subplot in the figure shows that the WIOB
  method agrees closely with the ZENO program when $\alpha=2$.

Recently, there has been great interest in developing a "fingerprint" of the shape of objects
through the calculation of the eigenvalues of the region based on solving the wave equation for objects of general shape, e.g.
Reuter et al. \cite{reuter:etal:2006} and
Peincke et al. \cite{peincke:etal:2007}. The eigenvalue spectrum is clearly advantageous to
other shape functionals as it involves the calculation of an infinite (large number) of numbers to specify object shape.
The $\alpha-$capacity likewise involves a ``spectrum'' of values that might better quantify object shape.
We note that the relative $\alpha-$capacity obeys the same isoperimetric property (see the Appendix),
i.e., the $\alpha$-capacity quantity is minimized by a sphere of all regions K of a fixed volume.
It is this isoperimetric property that makes the relatively capacity of interest in shape
classification and discrimination problems, as discussed for the 2-capacity and other shape
functionals having this fundamental isoperimetric property.

% -------------------------------------------------------------------------
\subsection{$\Capalpha$ of a thin ``coin''}

Here we consider a thin ``coin'' $K=\{ \boldx \in \reals^3 : x_2^2 + x_3^2 \le 1, -\epsilon \le x_1 \le \epsilon \}$.
Since $\epsilon$ will be small, $K$ is approximately a two dimensional set sitting inside of $\reals^3$.
This example illustrates two topics: using capacity to find Hausdorff dimension and subordination.

\begin{figure}
  \centering
  \includegraphics[width=\textwidth]{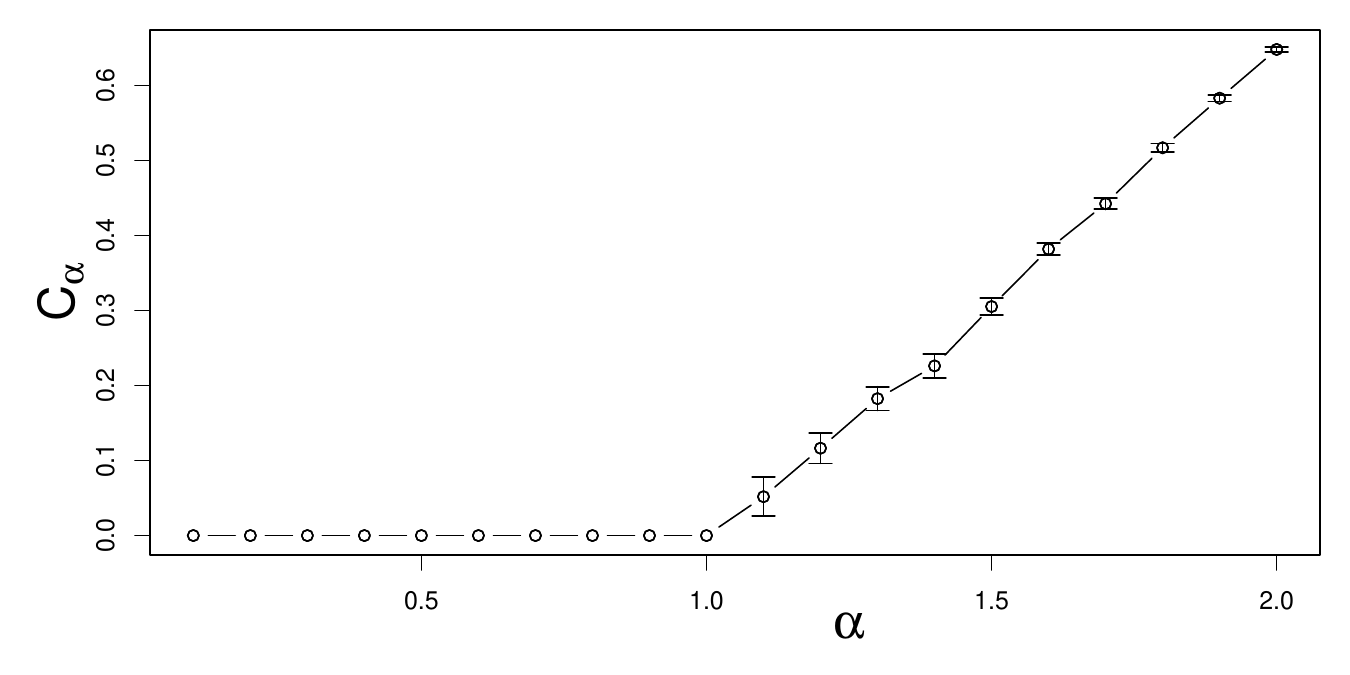}
  \caption{$\Capalpha$ of a thin ``coin'' with thickness $\epsilon= 0.01$, $n=10000$, $n_1=5000$, $p_\tau=0.99$, and confidence level $\delta=0. 95$.
  When $\alpha \le 1$, $\widehat{I}=\infty$, and confidence intervals for $\widehat{I}$, and therefore  for $\widehat{\Capalpha}$, are undefined.}\label{fig:coin_hausdorff_dim}
\end{figure}

For the first topic, we calculate the $\alpha$-capacity of the coin in three dimensions for a range of $\alpha$ values and see where the
capacity reaches 0.  Figure~\ref{fig:coin_hausdorff_dim} shows the result, where $\alpha=1$
defines a condition at which  the $\alpha$-capacity reaches 0.  This behavior arises because the Hausdorff dimension of the coin is $d-\alpha=3-1=2$.
(This conclusion follows from work by Taylor \cite{taylor:1955}, using a lemma of Frostman \cite{frostman:1935} for the calculation of Hausdorff dimension
of Brownian motion paths and other ``fractal'' sets. That work shows that the value of $\alpha$ where $\Capalpha(K)$ reaches 0  rigorously
defines the Hausdorff dimension of $K$.)  Correspondingly, the 2-capacity of the thin coin vanishes in four dimensions
 in the case of Brownian motion,
since the thin "coin" and Brownian motion have the same Hausdorff dimension, see Taylor \cite{taylor:1955},
Dvoretzky et al. \cite{dvoretzky:erdos:kakutani:1950}, and McKean \cite{mckean:1955}.

\begin{figure}
  \centering
  \includegraphics[width=\textwidth]{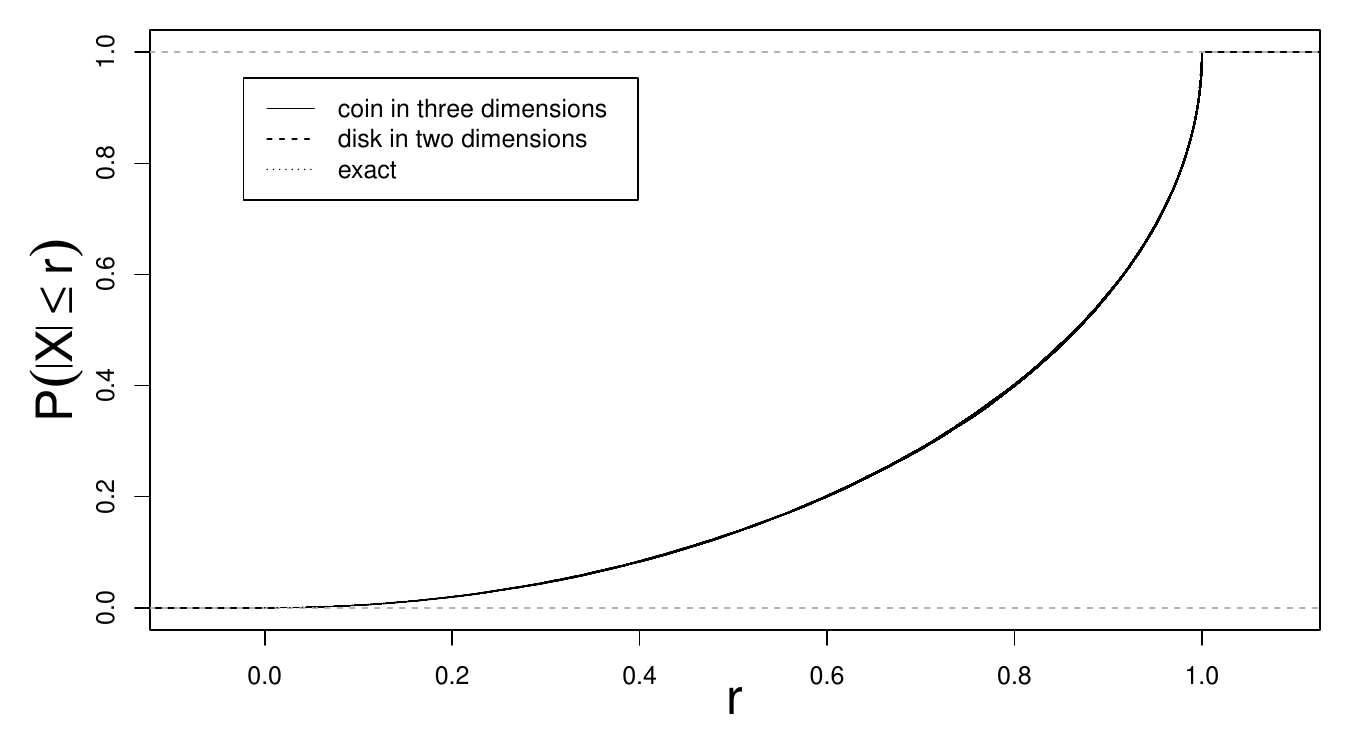}
  \caption{Cumulative distribution of radii of hitting distribution of a thin disk in $\reals^3$ with half thickness $\epsilon= 0.0001$,
  $\Rlaunch=5$, and $\Rescape=8$.
  The solid curve shows the cumulative of the $r_i$ values, the dashed curve shows the cumulative of the $s_i$ values, and
  the dotted curve shows the exact cumulative.  The three curves are visually indistinguishable}\label{fig:coin_hit_distrib}
\end{figure}

For the other topic, consideration of the hitting locations of Brownian motion ($\alpha=2$) in
$\reals^3$ on an infinitely thin ``coin'' provides an example of subordination, the reduction of Brownian
motion to a L\'{e}vy type process by ``decimating'' the  Brownian motion at random time points.  In particular,
the times at which the random walk in three dimensions hits a disc defines a random Cantor set of
time points of fractal dimension 1/2 (the subordinating set) and the path process connecting the
original random walk path positions at these time points defines a Cauchy process ($\alpha=1$)
 where the paths are then limited by construction to the unit disk in $\reals^2$.
We use the WOS method for Brownian motion to generate a sample hitting
distribution $\boldx_1,\ldots,\boldx_n \in \reals^3$ of the unit disk.  We then
%The last coordinate is dropped, giving points $\boldx^\prime_1,\ldots,\boldx^\prime_n \in \reals^2$.
perform a separate WIOB simulation for the unit disk in $\reals^2$ with an $\alpha=1$ path process and
determine hitting locations $\boldy_1,\ldots,\boldy_n \in \reals^2$.
To compare these to each other, the radial symmetry says it suffices to compare the distributions of univariate radii
$r_i=\sqrt{ x_{i2}^2+ x_{i3}^2 }$ (ignoring the small first coordinate) to univariate $s_i=\sqrt{ y_{i1}^2+ y_{i2}^2 }$.
Figure~\ref{fig:coin_hit_distrib} shows the cumulative distributions of these radii.  Also on the
plot is a third curve, which is  the exact cumulative distribution of the radii of the Cauchy equilibrium distribution derived
from (\ref{eq:equilibrium.dist}) with $d=2$ and $\alpha=1$.
% Mandelbrot, pg 291; Feller Vol II, pg 348?; Bochner (1955).
The view that the two dimensional Cauchy process is just a subordinated version of the three dimensional
Brownian path process, as described above, implies that the hitting distributions of the disk should then be the same.
We find that this might arise naturally when viewing the dynamics of a higher dimensional Brownian process in some space of reduced dimension.
We do not claim this is a universal explanation, but this may be a common mechanism.

% -------------------------------------------------------------------------
\subsection{Coverage probability for confidence intervals}

To assess the performance of the confidence interval method of Section~\ref{sec:conf.intervals},
we simulated $M=500$ runs with $K$ the unit ball in dimension $d=3$ and three $\alpha$ values.
The exact value of the $\alpha$-capacity is known in this case, so we counted how many of the 500 simulated 95\%
confidence intervals contained the exact value.
For $\alpha=0.5$, the observed coverage probability was 0.992; for
$\alpha=1.2$, the observed coverage probability was 0.966;  and for
$\alpha=1.9$, the observed coverage probability was 0.958.
The other parameters are $p_{\tau}=0.99$, $n=10000$, $n_1=n/2=5000$, % $n_2=500$,
$\Rlaunch=2$, $\Rescape=4$. The observed coverage probabilities are high, apparently due to
the large values of $v_1$.  It appears that even with the truncation at values of $w_{ij}$ above $\tau$,
the large values of $w_{ij}^*$ inflate the estimator $v_1$.  This will be investigated in future work.
Presently, observe that the confidence intervals are conservative, and that a larger simulation
can be used to get smaller confidence intervals.

% -------------------------------------------------------------------------
\section{Conclusion}

This work demonstrates that it is possible to precisely estimate  $\alpha$-capacity for
general sets $K$ in $d$-dimensions when $d > \alpha$.  In particular, it also novel in that it
estimates  standard electrostatic capacity in dimension $d > 3$.
Below, we first discuss the parts of the algorithm that are slow, and
then discuss some other applications of this approach.

The programs used here are written in R which is an interpreted
language, so would be faster if recoded in C/C+.  It is also straightforward to
parallelize the simulations of the hitting locations, because each path is independent of all others.
The run time is quicker if the object $K$ is centered near the origin and if the launch ball
is  the smallest centered ball that contains $K$.

Generating the hitting locations is currently slow for complicated objects, because
the R code computes the distance between the current location $\boldx$  and $K$ in step 2 of the WIOB method.
If $K$ is a polymer modeled by the union of thousands of beads, this is very slow.  The contribution of Juba et al. \cite{juba:keyrouz:mascagni:brady:2016}
was the development of an optimized data structure for $K$ that allows the efficient determination of
which beads are close to $\boldx$, greatly reducing the number of computations
required and a corresponding substantial reduction of the required computation time to obtain results having a prescribed uncertainty.
This technique can be applied here to speed up simulations.

The time consuming part of the $\alpha$-capacity estimation is
the computation of $\widehat{I}_1$, which is quadratic in $n_1$.  Calculation of $v_1$ is of order $n_1$.
The algorithm is also memory intensive, requiring
$n^2$ memory locations for the matrix $[ w_{ij} ]_{i,j=1}^n$.  It is possible to
improve on this and if time is not a concern, one can keep just the $n$ hitting locations
and compute entries of the $W$ matrix as needed.

This approach can be used to give an approximate solution to fractional diffusions,
see Kyprianou et al. \cite{kyprianou:osojnik:shardlow:2018}.
In this context, we have described the steps needed to solve the exterior problem,
i.e. starting a fractional diffusion outside of $K$ and running the path until it hits
$K$.  These same tools can be adapted to solve the interior problem where we start
inside $K$ and run the process until it leaves the set using the WIOB method.
Hunt et al. \cite{hunt:douglas:bernal:1995} discuss a potentially important application of this type of calculation to
modeling the velocity field of high Reynolds number pipe flow.

Finally, because the parameter $\alpha$ can be varied in the interval $(0,2]$, we
obtain a ``spectrum'' of capacity estimates for any set $K$.  Dividing these $\alpha$-capacity
values by the capacity of a ball having the same volume as $K$ gives shape information.
Such a ``spectrum'' should be useful in shape discrimination.
These values can provided information about the interior of sets, as in the
cases of hollow squares considered in Section~\ref{sec:solid.hollow.cubes}.
In the case of fractal sets, it should be possible to estimate the Hausdorff dimension by varying $\alpha$ and finding the value
at which $\Capalpha=0$.  This approach has long been used in analytic estimates of fractal sets such as
Brownian motion through the use of upper and lower bounds as in Taylor \cite{taylor:1961}, but this method
has apparently never been implemented numerically using path integration methods.  We expect
this method of calculating fractal dimension to be quite general and capable of precise dimension estimates.

% -------------------------------------------------------------------------
\appendix
\section{Properties of capacity and stable processes}

In this section we collect technical results on stable processes and capacity that are used above.
We do not prove these results, only give references to where these facts can be found.
General information on stable distributions can be found in Nolan \cite{nolan:book1} and information
  on stable processes can be found in Samorodnitsky and Taqqu \cite{samorodnitsky:taqqu:1994},
  Sato \cite{k.sato:1999}, and Bogdan et al. \cite{bogdan:byczkowski:kulczycki:ryznar:song:vondracek:2009}.
 Landkof \cite{landkof:1972} gives the basic properties of $\alpha$-capacity in Chapter~2.
% Here we state the properties that are used in this paper.

\bigskip
\begin{description}
  \item[Translation and scaling.] If $K$ is any compact set, any scale $r > 0$, and shift $\boldx$,
   \begin{equation}\label{eq:translation.n.scaling}
   \Capalpha(r K + \boldx) = r^{d-\alpha} \Capalpha(K).
   \end{equation}
   Note that $\Capalpha(r K) = r \Capalpha(K)$ in only two cases:
   electrostatic capacity ($\alpha=2$) when $d=3$ or Cauchy capacity ($\alpha=1$) when $d=2$.

  \item[Transcience/recurrence.]  For $0 < \alpha \le 2$, an $\alpha$-stable motion is
   recurrent if $d \le \alpha$, transient if $d > \alpha$.  See Sato \cite{k.sato:1999}, Chapter 7.
   The method described here for calculating $\alpha$-capacity can only be used
   in the transient case.

  \item[$\alpha$-capacity of a ball.] Let $\ball \subset \reals^d$ be the unit ball.  Then for any shift $\boldx$ and scale $r>0$
  \begin{equation}\label{eq:ball_capacity}
   \Capalpha(\boldx + r \ball ) = r^{d-\alpha} \Capalpha( \ball ) = r^{d-\alpha}
   \frac{ \Gamma(d/2)}{ \Gamma(\alpha/2) \Gamma((d-\alpha+2)/2)}.
 \end{equation}

   This is given on page 111 of Takeuchi \cite{takeuchi:1964}.

  \item[Isoperimetric and symmetrization properties] For all regions of fixed volume and for all $0 < \alpha \le 2$,
    the $\alpha-$capacity is     minimized for a sphere. This was first proved by  Watanabe \cite{watanabe:1983}.
    We may more generally expect objects having greater symmetry to have a lower $\alpha-$capacity. See
      Betsakos \cite{betsakos:2004a}, \cite{betsakos:2004b}, and    Mendez-Hernandez \cite{mendezhernandez:2006}Kojar \cite{kojar:2015}
      for proofs and discussions of these facts.

  \item[Subadditivity.] For compact Borel sets $A, B \subset \reals^d$,
   $\Capalpha(A \cup B) \le \Capalpha(A) + \Capalpha(B)$.  See \cite{landkof:1972}, Chapter 2.  % page 141

  \item[Hitting probability for a ball.] A stable motion $\boldX_t$ hits the ball $\boldx+r \ball$ if and
  only if the shifted and scaled process $(\boldX_t-\boldx)/r$ hits the unit ball $\ball$.
  Likewise, $\boldX_t$ leaves the ball $\boldx+r \ball$ if and only the shifted and scaled process $(\boldX_t-\boldx)/r$
  leaves the unit ball $\ball$.
  Because of this, we can restrict ourselves to considering entering and leaving the unit ball.
  For the unit ball and starting at $\boldx$ with $|\boldx| > 1$,
  \begin{equation}\label{eq:hit.ball.prob.gen.alpha}
  P(\boldX_t \text{ hits } \ball \, | \, \boldX_0 = \boldx ) =
  \begin{cases} 1 & \boldX_t \text{ is recurrent } \\
      F\left(\frac{1}{|\boldx|^2};\frac{d-\alpha}{2}, \frac{\alpha}{2} \right) &  \boldX_t \text{ is transient},
  \end{cases}
  \end{equation}
  where $F(\cdot;a,b)$ is the cumulative distribution function of a Beta distribution with parameters
  $a$ and $b$.  This follows from a change of variables in Corollary~2 of Blumenthal et al. \cite{blumenthal:getoor:ray:1961}.

   This can also be expressed as $P(\boldX_t \text{ hits } \ball \, | \, \boldX_0 = \boldx ) = P(T_\ball < \infty)$, where
   $T_\ball = \inf_{t > 0} \{ \boldX_t \in \ball \, | \, \boldX_0 = \boldx )$ is the first hitting time.
       A series expansion for the exact probability of hitting the unit ball is
   $$P(T_\ball < \infty| \,|x|=r ) =  \sum_{j=0}^\infty c_j r^{-(d-\alpha+2j)},$$
   where the coefficients are given by
   $c_0 = \Gamma(d/2)/( \Gamma( (d-\alpha+2)/2) \Gamma(\alpha/2) )=\Capalpha(\ball)$ and recursively for $j > 0$ by
   $$c_j = \frac{( 2j - \alpha)(d-\alpha+2(j-1))}{ 2j(d-\alpha+2 j) } c_{j-1}.$$
   This follows from the Taylor series expansion for the incomplete Beta function around 0.
   If $\alpha=2$ and $d \ge 2$, $c_0=1$ and all other coefficients are 0, so
   $P(T_\ball < \infty| \,|x|=r ) = r^{-(d-\alpha)}$.  When $ 0 < \alpha < 2$,
   some algebra shows all $c_j > 0$ and they are strictly decreasing,
   so a truncated series is always a lower approximation for the probability of hitting a ball.

   For a ball of arbitrary center $\boldy$ and radius $r > 0$, when $\boldx$ is not in the ball,
   scaling shows that in the transient case
     $$P(\boldX_t \text{ hits } \boldy + r \ball \, | \, \boldX_0 = \boldx ) =
       F(r^2/|\boldx-\boldy|^2;(d-\alpha)/2, \alpha/2).$$

   \item[Hitting location for a ball.] In the Brownian case,  the distribution of the hitting location starting from $\boldx$ outside
   the unit ball is concentrated on the surface of the sphere.  When $d=1$, the hitting location for the ball
   is the nearest endpoint.  When $d > 1$, it has density given by the Poisson kernel:
   $$ f_{\mathrm{Poisson}}(\boldx,\boldy) = \frac{\left|1-|\boldx|^2 \right| } {|\boldx - \boldy|^d}, $$
   see M\"orters and Peres \cite{morters:peres:2010}, Theorem 3.44.
   When $0 < \alpha < 2$, Blumenthal et al. \cite{blumenthal:getoor:ray:1961} showed that
   an $\alpha-$stable motion starting at $\boldx$ outside the unit ball has hitting density for $|\boldy|<1$
   $$f_{\mathrm{BGR}}(\boldx,\boldy) =
   \begin{cases}  \displaystyle{\frac{c(\alpha,d) \left|1-|\boldx|^2 \right|^{\alpha/2}  }
                                   {\left|1-|\boldy|^2 \right|^{\alpha/2} |\boldx - \boldy|^d}}
                                           & \alpha < d \mbox{ or } d= \alpha = 1 \\
                                     \displaystyle{  \displaystyle{\frac{c(\alpha,1) \left|1-x^2 \right|^{\alpha/2}  } {\left|1-y^2 \right|^{\alpha/2} |x - y|^d}} } +\frac{c(\alpha,1)(\alpha-1) }{|1-y^2|^{\alpha/2}}   \times &  \\
                                      \hspace{10mm} \displaystyle{  \int_1^{|\boldx|} (u^2-1)^{\alpha/2-1} du} & d=1,  1 < \alpha < 2
                         \end{cases}$$
   where $c(\alpha,d) = \pi^{-((d/2)+1)} \Gamma(d/2) \sin( \pi \alpha /2 ) $.  For a ball of center $\boldx_0$ and radius $r$, shifting and scaling the above formulas
   can be used.

   Devroye and Nolan \cite{devroye:nolan:2022} show how to efficiently simulate from this density.
   That method assumes that $\boldx$ is on the $x_1$ axis for simplicity.
   To handle a general starting location $\boldx$, see below.

   \item[Hitting location for the complement of a ball.]  In part of the WIOB method, an $\alpha-$stable process is started at
   a point $\boldx$ on the interior of a ball and it is run until it exits the ball to a point $\boldy$.  Because of the shifting and scaling
   properties, it suffices to consider the centered unit ball.

   When $\boldx=0$, use $\boldy = \boldZ / \sqrt{B}$, where $\boldZ$ is uniform on the unit sphere and
   $B \sim $Beta($\alpha/2,1-\alpha/2$).
    When $\boldx \ne 0$, duality can be used.  \cite{bogdan:zak:2006} call this the Kelvin transform.
    Reflect the point $\boldx$ across the boundary sphere: $\boldx \to \boldx^\prime=\boldx/|\boldx|^2 $.
   Since $\boldy$ is outside the ball, we can run a stable process starting at $\boldx^\prime$ until it hits the ball at some point
   $\boldy^\prime$.  Then reflect $\boldy^\prime$ to get $\boldy = \boldy^\prime/ |\boldy^\prime|$ to get the desired random vector.

  \item[Generating vectors uniformly distributed on $\sphere$.] If $\boldZ = (Z_1,\ldots,Z_d)$ has
  independent $N(0,1)$ components, $\boldY = \boldZ/|\boldZ|$ is uniformly distributed on the
  unit sphere $\sphere$.

  \item[Generating isotropic stable vectors.]  Suppose $\gamma > 0$, $0 < \alpha < 2$,  and $S > 0$ is a positive
  $(\alpha/2)$-stable random variable with scale $\gamma \cos(\pi \alpha/4)^{1/\alpha}$ and zero shift.
  This can be directly simulated using the method in Chambers et al. \cite{chambers:mallows:stuck:1976}.  In the $\alpha<1,\beta=1$ case the formula is
  $$S = \frac{ \gamma \sin(\alpha(\pi/2 + \theta)) \cos(\alpha \pi/2 + (\alpha-1) \theta) } {\cos(\pi \alpha/2) \cos(\theta)^{1/\alpha} W ^{(1-\alpha)/\alpha} } ,  $$
  where $\theta$ is uniformly distributed on $(-\pi/2,\pi/2)$ and $W$ is an independent exponential r.v. with mean 1.
  If $\boldZ = (Z_1,\ldots,Z_d)$ has  independent $N(0,1)$ components, then
 $\boldY = S^{1/2} \boldZ$ is an isotropic stable random vector with characteristic function
 (\ref{eq:isotropic.stable.step}).

 \item[Equilibrium measure for a ball.]  When $0 < \alpha < 2$,  Takeuchi \cite{takeuchi:1964}, page 111,
gives the density of the equilibrium distribution for the unit ball as
 \begin{equation}
 f(\boldy) = \frac{\sin(\pi \alpha/2) \Gamma(d/2)}{\pi^{1 + d/2}} \left(1- |\boldy|^2  \right)^{-\alpha/2}. \label{eq:equilibrium.dist}
 \end{equation}
This can be simulated by taking $\boldZ=(Z_1,\ldots,Z_d)$ where the $Z_i$ are i.i.d. $N(0,1)$ and
computing
$$ \boldY = \sqrt{B} \boldZ/|\boldZ|,$$
where $B \sim $Beta($d/2,1-\alpha/2$).

When $\alpha=2$, the equilibrium measure is uniform on the
 sphere, which can be simulated using the method above: $\boldY = \boldZ/|\boldZ|$.

 % info on Given's rotations in Golub and Van Loan, 3rd Edition, pg 215-218
  \item[Arbitrary starting locations]  The methods described in Devroye and Nolan \cite{devroye:nolan:2022} for
  simulating hitting a ball assume the stable motion starts at a point $(\lambda,0,\ldots,0)$ on
  the $x_1$-axis.  For a general starting position $\boldx$ find a $d \times d$ rotation matrix $U$ such
  that $\boldx^\prime = U\boldx=(|\boldx|,0,\ldots,0)$, run the algorithm with $\lambda=|\boldx|$ to get a hitting
  location $\boldy^\prime$, and rotate back to get $\boldy = U^{-1} \boldy^\prime$.  One fast
  way to find such a rotation is to use a sequence of Given's rotations to successively zero
  out the $d$-th component of $\boldx$, then the $(d-1)$-th component, etc.  Then apply
  these rotations in reverse to get $U^{-1}$.  Since $\boldx^\prime$ can be computed by
  just calculating $|\boldx|$, $U$ is not needed for this algorithm, we only need to find $U^{-1}$.

  The algorithms to compute $U$ and $U^{-1}$ only take $\mathcal{O}(d)$ steps since the matrices are sparse.
  A third function is provided to compute a rotation matrix that takes any non-zero vector $\boldx$ to any other non-zero vector $\boldy$
  by composing a rotation from an original vector $\boldx$  to the $x_1$-axis, then a rotation from the $x_1$-axis
  to the desired $\boldy$ direction.

  \item[More information from simulations]  The code implements hitting locations in both the simple random
  walk method of Section~\ref{sec:simple.rw} and WIOB method of Section~\ref{sec:WIOB}.  In this paper,
  we have only used the final hitting location of a stable motion, but the program gives the option of
  saving the path, which may be of interest in solving other problems, e.g. fractional diffusions.

\end{description}

% -------------------------------------------------------------------------

\section*{Acknowledgments}
The authors would like to thank David Gerard of American University for discussions about U-statistics,
Renming Song of the University of Illinois for alerting us to Theorem~2 in \cite{port:1967}, and two
reviewers whose comments improved the paper.

%\bibliographystyle{siamplain}
%\bibliography{AlphaCapacity}

% -------------------------------------------------------------------------
\end{document}